\documentclass[%
 aip,%
 cp,%
 amsmath,amssymb,%
 reprint,%
 groupedaddress,
]{revtex4-1}

\usepackage{graphicx}
\usepackage{dcolumn}
\usepackage{bm}
\usepackage{subfigure}
\usepackage{float}
\usepackage{xcolor}
\usepackage{epstopdf}

\begin{document}

\preprint{AIP/123-QED}

\title[]{Explicit Filtering in Large Eddy Simulation of Barotropic Turbulence in Spectral Space}

\author{Leila~N. Azadani}
\email{leila@vt.edu.}
\author{Anne~E. Staples}%
 \affiliation{ 
Department of Engineering Science and Mechanics, Virginia Tech, Blacksburg, VA 24061, USA.
}%


\begin{abstract}
Explicit filtering in large eddy simulation (LES) of a turbulent barotropic flow on the sphere in spectral space is studied and compared to implicit filtering. Here, a smooth filter is applied to the nondivergent barotropic vorticity equation (BVE) on the rotating sphere to divide the flow field into resolved scale (RS) and subfilter scale (SFS) motions. A portion of the SFS motions are reconstructed theoretically using the approximate deconvolution model (ADM). While the unreconstructed portion consists of the subgrid scale (SGS) motions and needs to be modeled separately. In order to investigate the effects of the explicit filtering alone no SGS model is used. It is shown that the explicit filtering accurately tracks the evolution of the coherent structures in two-dimensional turbulent flow on the rotating sphere, whereas the implicit filtering does not. It is also shown that explicit filtering improves the results of the temporal variation of the total kinetic energy and the total enstrophy and the variation of the energy spectrum with wavenumber compare to the implicit filtering. Although explicit filtering is more expensive than implicit filtering it increases the accuracy of the computations and improves the results, particularly where the location of coherent structures is concerned, a topic of particular importance in LES of atmospheric flows for climate and weather applications.
\end{abstract}

\maketitle

\section{Introduction}
\label{sec:intro}
Large eddy simulation (LES) is a powerful technique to study turbulent flows. In LES, length scales smaller than a certain cutoff scale are eliminated from the large-scale flow motions by applying a proper low-pass filter. The filtering operation in LES can be implicit or explicit. In implicit filtering the computational grid and discretization schemes are considered to be the low-pass filter. In this case the flow field is divided into the large resolved scales and the small subgrid scales. In explicit filtering, though, in addition to the implicit filtering due to the computational grid and discretization schemes, an explicit filter with a width larger than the mesh spacing is applied and separates the flow field into resolved scales and subfilter scales. The subfilter scales are again divided into resolvable subfilter scale (RSFS) and unresolvable subfilter scale (URSFS) motions. Resolvable subfilter scales can be reconstructed theoretically from the resolved flow field, while unresolvable subfilter scales, which are traditionally called subgrid scales, need to be modeled separately \cite{Carati, Zhou}. 

Implicit filtering is the most commonly used technique in LES of turbulent flows because it is computationally less expensive and less complicated than explicit filtering. However, implicit filtering is associated with some numerical issues. First of all, in implicit filtering the derivative operation acts as a low-pass filter in the spatial direction in which the derivative is taken so each term in the Navier-Stokes equations is subjected to a different one-dimensional filter and it is not possible to derive the LES equations through applying a single three-dimensional filter. In addition, controlling the frequency content of the advective term is not possible in implicit filtering. Because of nonlinear interactions of turbulent motions high frequency motions are generated and contaminate the resolved scales of the flow. Although a suitable SGS model should consider this effect, such a constraint has rarely been included. The final shortcoming of implicit filtering is the inability to control truncation errors due to discretization schemes. The truncation error is small for high-order numerical methods and increases with decreasing accuracy of the discretization scheme. These errors interfere with the dynamics of the smallest resolved scales. Since these scales have an important role in modeling SGS motions, they need to be computed accurately \cite{Lund}. Explicit filtering overcomes some of the difficulties associated with the implicit filtering. However, explicit filtering reduces the effective grid resolution and increases the computational cost. Furthermore, in explicit filtering the filtering operator and the differentiation operator need to commute, otherwise the commutation error should be considered when solving the Navier-Stokes equations. Choosing between implicit and explicit filtering depends on the desired accuracy and cost of the computation. 

The concept of explicit filtering is mixed with reconstruction models. Reconstruction models are used to account for the behavior of the RSFS motions. The first reconstruction model dates back to Leonard \cite{Leonard}. He provided an analytical expression based on Taylor series expansions of the filtering operator to reconstruct the filtered scales due to explicit filtering. Clark \textit{et al.} \cite{Clark} improved the method proposed by Leonard and introduced the gradient or nonlinear or tensor-diffusivity model. 

The scale similarity model of Bardina \cite{Bardina} is the other popular model for reconstructing the RSFS motions. In the scale similarity model it is assumed that the smallest resolved scales are similar to the largest unresolved scales. Thus, the unknown unfiltered quantities can be approximated by the filtered quantities. 

Both gradient and scale similarity models show good correlation in a priori tests while in a posteriori tests they do not dissipate enough energy. These models are just used to reconstruct the RSFS motions and need to be combined with an SGS model. This combination introduces so-called mixed models where the base model, either gradient or scale similarity, is combined with an eddy-viscosity model to ensure enough subgrid scale dissipation. 

Clark \textit{et al.} \cite{Clark} and Winckelmans and Jeanmart \cite{Winckelmans1} combined the gradient model with the traditional Smagorinsky model \cite{Smagorinsky} to simulate decaying isotropic turbulence. 

Bardina \textit{et al.} \cite{Bardina} combined the scale similarity model with the Smagorinsky model and tested the method against experimental results on homogeneous isotropic, rotating, and shear turbulence. 

Zang \textit{et al.}\cite{Zang} mixed the scale similarity model with the dynamic Smagorinsky model \cite{Germano} to study turbulent flows in a lid-driven cavity. They noticed that the results obtained using the mixed model show better agreement with experiments compare to those obtained using dynamic Smagorinsky.

Winckelmans \textit{et al.} \cite{Winckelmans2} coupled the tensor-diffusivity model with the dynamic Smagorinsky model. They applied the method to the simulation of decaying isotropic turbulence and turbulent channel flow.

Shah and Ferzige \cite{Shah} showed that including higher order terms in approximations of the unfiltered quantities in terms of filtered ones in the scale similarity model provides sufficient dissipation. They applied their method to simulations of channel flow and compared the results with the results obtained from Smagorinsky and dynamic Smagorinsky models.

The velocity estimation model of Domaradzki  and Saiki \cite{Domaradzki} is another attempt to reconstruct the unfiltered flow variables from the filtered variables. Domaradzki and Saiki estimated the unfiltered velocity field by expanding the resolved velocity field to subgrid scales two times smaller than the grid scale. This model was applied to the computation of turbulent channel flow and the results were compared with the classical Smagorinsky model. 

The approximate deconvolution model (ADM) of Stolz \cite{Stolz1} aims at approximating the unfiltered quantities based on repeated application of an inverse filter to the filtered quantities. This model has been successfully applied to computations of incompressible \cite{Stolz2} and compressible \cite{Stolz3} turbulent flows. 

Chow \textit{et al.} \cite{Chow} used ADM to reconstruct the resolvable subfilter scales and the dynamic Smagorinsky model to model the effects of subgrid scales in the computation of the atmospheric boundary layer. Their results showed significant improvements in accuracy of the results compare to the results obtained using implicit filtering and a dynamic eddy-viscosity model. 

San \textit{et al.} \cite{San1} applied ADM to the computation of two-dimensional barotropic oceanic flows. Results obtained using explicit filtering and ADM showed the correct four-gyre circulation structure predicted by direct numerical simulation (DNS) results while applying the implicit filtering yielded a two-gyre structure, which is not consistent with the DNS data. San \textit{et al.} \cite{San2} also applied ADM in simulation of a stratified two-layer ocean model. 

 Following San \textit{et al.} we would like to study the effects of explicit filtering in computations of two-dimensional barotropic atmospheric flows. San \textit{et al.} \cite{San1} ignored the effects of the Earth's curvature in their computations, while this effect cannot be neglected in simulation of large-scale atmospheric and oceanic circulations \cite{Jacobson}. We consider this effect by solving the governing equations in spherical coordinates. Computations of geophysical flows in spherical coordinates are best  performed by using spectral methods because these methods are accurate and account for the spherical geometry of the Earth. Here, we use a spectral method based on spherical harmonic transforms to solve the barotropic vorticity equation (BVE).

Conventional wisdom dictates that explicit filtering in spectral space leads to no improvement over implicit filtering. Although spectral simulations of isotropic turbulence by Winckelmans \textit{et al.} \cite{Winckelmans2} verified this wisdom, spectral computations of turbulent channel flow by  Domaradzki and Saiki \cite{Domaradzki} and Stolz \textit{et al.} \cite{Stolz2} showed significant improvement in the explicit filtering results over implicit filtering results. There are two main differences between the work by Winckelmans \textit{et al.} \cite{Winckelmans2} and the studies by Domaradzki and Saiki \cite{Domaradzki} and Stolz \textit{et al.} \cite{Stolz2}. First, the nature of the problem is different in these two cases and obviously different spectral methods were employed. Spectral computations of isotropic turbulence are usually performed based on Fourier transforms while in spectral simulations of turbulent channel flow a combination of Fourier and Chebyshev transforms is applied. Secondly, these people have used different subfilter reconstruction models. Winckelmans \textit{et al.} \cite{Winckelmans2} used the tensor-diffusivity model to reconstruct the filtered-scale tensor, Domaradzki and Saiki \cite{Domaradzki} applied the velocity estimate method to reconstruct the unfiltered velocity field from the filtered velocity field, and  Stolz \textit{et al.} \cite{Stolz2} used the ADM for reconstructing the subfilter scales. Therefore, it is unclear whether the spectral basis functions or subfilter reconstruction models led to the improved results with explicit filtering.

 Here, we use a different spectral method based on spherical harmonic transforms which are a combination of Fourier transforms and Legendre transforms. In order to reconstruct the unfiltered flow variables we use both exact deconvolution by applying the exact inverse filter to the filtered flow field and the approximate deconvolution model. The goal of this paper is to investigate the effectiveness of explicit filtering using a spherical harmonics spectral method and a high-order reconstruction model.

The organization of this paper is as follows; the governing equations are presented in section~\ref{sec:gov}. ADM is introduced in section ~\ref{sec:adm}. In section ~\ref{sec:nm} the numerical method is discussed. The results and discussion are given in section~\ref{sec:res} and conclusions are made in section ~\ref{sec:con}.

\section{Governing equations}
\label{sec:gov}
The barotropic vorticity equation (BVE) is the simplest nontrivial model to study dynamics of large-scale motions of planetary atmospheres and oceans. Charney \textit{et al.} \cite{Charney} performed the first successful numerical weather prediction based on the BVE. The BVE describes the motion of a two-dimensional, nondivergent, incompressible fluid on the rotating sphere and is given by
\begin{equation}
\frac{\partial\zeta}{\partial t}+J(\psi, \zeta+f)=(-1)^{p+1}\nu_{2p}\nabla^{2p}\zeta
\label{eq:bve}
\end{equation}
\begin{equation}
\zeta=\nabla^2\psi
\label{eq:laplas}
\end{equation}
where $\zeta(\lambda, \mu,t)$ is the vertical component of the vorticity, $\psi(\lambda, \mu, t)$ is the streamfunction, $f=2\Omega sin\theta$ is the Coriolis parameter, $\nu_{2p}$ is the hyperviscosity coefficient and $J$ is the horizontal Jacobian operator on the sphere, and is defined as
\begin{equation}
J(\psi, \zeta+f)=\frac{1}{R^2}\left [\frac{\partial\psi}{\partial\lambda}\frac{\partial(\zeta+f)}{\partial\mu}-\frac{\partial\psi}{\partial\mu}\frac{\partial(\zeta+f)}{\partial\lambda}\right]
\end{equation}
where $R$ is radius of the sphere, $-\pi\leqslant \lambda \leqslant \pi$ and $-\pi/2\leqslant \theta \leqslant \pi/2$ are the longitude and latitude, $\mu=sin\theta$, and $\Omega$ is the rotation rate of the sphere.

Eq.~(\ref{eq:bve}) is nondimensionalized by taking the radius of the sphere, $R$, as the length scale, $U$ as the characteristic velocity scale and $R/U$ as the advection time scale. Nondimensionalizing Eq.~(\ref{eq:bve}) introduces the Rossby number, an important physical parameter in a rotating system, which is defined as
\begin{equation}
Ro=\frac{U}{2R\Omega}
\label{eq:ro}
\end{equation}
and is said to be the ratio of the inertial force to the Coriolis force.

The BVE is solved under periodic boundary conditions in the $\lambda$ direction. In the $\mu$ direction $\zeta$ should be independent of $\mu$ on the poles so
\\* $$\zeta(\lambda, \mu, t)=\zeta(\lambda+2\pi, \mu, t) $$
and
\\* $$ \zeta(\lambda, -1,t) \text{ and } \zeta(\lambda, 1, t) \text{ independent of } \lambda$$

The initial conditions we use are based on the following initial energy spectrum \cite{Cho}
\begin{equation}
E(n,0)=\frac {An^{\gamma/2}}{(n+n_0)^\gamma}
\label{eq:e0}
\end{equation}
where $A$ is a normalization constant, $n_0$ is the peak wavenumber of the energy spectrum, and $\gamma$ is used to control the width of the spectrum. Here, $n_0$ and $\gamma$ are set to 10 and 20, respectively.

\section{Approximate deconvolution and exact deconvolution}
\label{sec:adm}
In LES, a spatial filter is applied to the fluid field to separate flow motions into large and small scales. The filtering operation is mathematically represented as a convolution product
\begin{equation}
\bar u=G\ast u
\label{eq:conv}
\end{equation}
where $u$ is a typical flow variable, $\bar u$ is the filtered variable, $G$ is the filter kernel, and the subfilter flow variable, $u'$, is defined as
\begin{equation}
u'=(1-G)\ast u.
\label{eq:convu}
\end{equation}
Approximate deconvolution is a process for approximating the unfiltered variable by applying an inverse filter
\begin{equation}
u=G^{-1}\ast \bar u.
\label{eq:deconv}
\end{equation}
In Eq.~(\ref{eq:deconv}), $G^{-1}$ is the inverse filter and can be defined by the Neumann series
\begin{equation}
G^{-1}=\sum_{i=1}^{\infty}(I-G)^{i-1}
\label{eq:neu}
\end{equation}
where $I$ is the identity operator. This nonconvergent Neumann series can be approximated by the Van Cittert equation as
\begin{equation}
D_N\approx G^{-1}=\sum_{i=1}^{N}(I-G)^{i-1}
\label{eq:van}
\end{equation}
where,
\begin{eqnarray}
D_1&=&I, \nonumber\\
D_2&=&2I-G,  \nonumber\\
D_3&=&3I-3G+GG,  \nonumber\\
D_4&=&4I-6G+4GG-GGG,  \nonumber\\
D_5&=&5I-10G+10GG-5GGG+GGGG,  \nonumber\\
\vdots
\end{eqnarray}
An approximate deconvolution of $\zeta$ can now be obtained as
\begin{equation}
\zeta\approx \zeta^\ast=D_N\bar{\zeta}.
\label{eq:van}
\end{equation}
 If we choose $N=3$, $\zeta$ and $\psi $ can be approximated as
\begin{equation}
\zeta\approx \zeta^\ast=3\bar{\zeta}-3\bar{\bar{\zeta}}+\bar{\bar{ \bar {\zeta}}}
\label{eq:adq}
\end{equation}
\begin{equation}
\psi\approx \psi^\ast=3\bar{\psi}-3\bar{\bar{\psi}}+\bar{\bar{ \bar {\psi}}}.
\label{eq:adp}
\end{equation}
Using Eqs.~(\ref{eq:adq}) and (\ref{eq:adp}) the nonlinear Jacobian term can be approximated as
\begin{equation}
\overline{J(\psi,\zeta)}\approx \overline {J(\psi^\ast,\zeta^\ast)}.
\label{eq:j}
\end{equation}
Additionally, in spectral methods, the unfiltered flow fields can be exactly reconstructed by applying the exact inverse filter to the filtered fields so the exact deconvolution of filtered flow variables is given that
\begin{equation}
\zeta= \zeta^\ast=\bar{\zeta}/G
\label{eq:efz}
\end{equation}
\begin{equation}
\psi= \psi^\ast=\bar{\psi}/G
\label{eq:efp}
\end{equation}
and
\begin{equation}
\overline{J(\psi,\zeta)}=\overline {J(\psi^\ast,\zeta^\ast)}.
\label{eq:je}
\end{equation}
We use both methods here to keep the presentation as general as possible. Substituting Eq.~(\ref{eq:j}) from the approximate deconvolution model or Eq.~(\ref{eq:je}) from exact deconvolution, and $f=2\Omega\mu$ into Eq.~(\ref{eq:bve}) the filtered barotropic vorticity equation is given by
\begin{equation}
\frac{\partial \bar\zeta}{\partial t}+J(\bar\psi, \bar\zeta)+\frac{2\Omega}{R^2}\frac{ \partial\bar\psi}{\partial \lambda}=(-1)^{p+1}\nu_{2p}\nabla^{2p}\bar\zeta + S^\ast
\label{eq:bvef}
\end{equation}
where $S^\ast$ is the subfilter scale tensor and is given by
\begin{equation}
S^{\ast}=J(\bar\psi, \bar\zeta)-\overline {J(\psi^*,\zeta^*)}.
\label{eq:sfs}
\end{equation}
To perform explicit filtering in LES of the BVE we need to define an appropriate filter. We use the differential filter, which in physical space is defined as
\begin{equation}
\bar u-\frac{\partial}{\partial x_j}\left(\alpha \frac{\partial \bar u}{\partial x_i}\right)=u
\label{eq:dif}
\end{equation}
where $\alpha=\alpha(x)$ is related to the filter width. We use the differential filter with a constant width, so Eq.~(\ref{eq:dif}) becomes
\begin{equation}
\bar u-\alpha\nabla^2\bar u=u.
\label{eq:difc}
\end{equation}
The differential filter kernel in spectral space will be given in section \ref{sec:nm}.

\section{Numerical method}
\label{sec:nm}
Eq.~(\ref{eq:bvef}) is solved in spectral space by expanding each variable into a series of spherical harmonics as
\begin{equation}
\zeta(\lambda, \mu, t)=\sum_{m=-\mathcal{N}}^{\mathcal{N}}\sum_{n=|m|}^{\mathcal{N}}\zeta_{n}^{m}(t)Y_{n}^{m}(\lambda, \mu)
\label{eq:sp}
\end{equation}
where $\zeta_{n}^{m}(t)$ are the complex spectral coefficients of $\zeta(\lambda, \mu, t)$, $n$ is the total wavenumber, $m$ is the zonal wavenumber, $\mathcal{N}$ denotes the truncation wavenumber and $Y_{n}^{m}$ are spherical harmonics defined by
\begin{equation}
Y_{n}^{m}(\lambda, \mu)=P_{n}^{m}(\mu)e^{im\lambda}
\label{eq:sh}
\end{equation}
where $P_{n}^{m}(\mu)$ are the normalized associated Legendre polynomials and $i=\sqrt{-1}$. Eq.~(\ref{eq:sp}) is truncated using triangular truncation which, unlike rhomboidal truncation, is rotationally symmetric. The nonlinear Jacobian term in Eq.~(\ref{eq:bvef}) is computed using the pseudospectral method, in which the nonlinear term is calculated in physical space and transformed to spectral space. 

The energy and enstrophy spectra are defined as
\begin{equation}
E(n,t)=\frac{1}{2}\sum_{m=-n}^{n}\frac{R^2}{n(n+1)}|\zeta_{n}^{m}(t)|^2
\label{eq:es}
\end{equation}
\begin{equation}
Ens(n,t)=\sum_{m=-n}^{n}n(n+1)E(n,t)
\label{eq:ens}
\end{equation}
and the total kinetic energy and enstrophy are given by
\begin{equation}
E(t)=\sum_{n=0}^{\mathcal{N}}E(n,t)
\label{eq:es}
\end{equation}
\begin{equation}
Ens(t)=\sum_{n=0}^{\mathcal{N}}Ens(n,t).
\label{eq:ens}
\end{equation}
\begin{figure}[!b] 
\begin{center}
\includegraphics[trim =1mm 0mm 20mm 20mm, clip, width=0.4\textwidth]{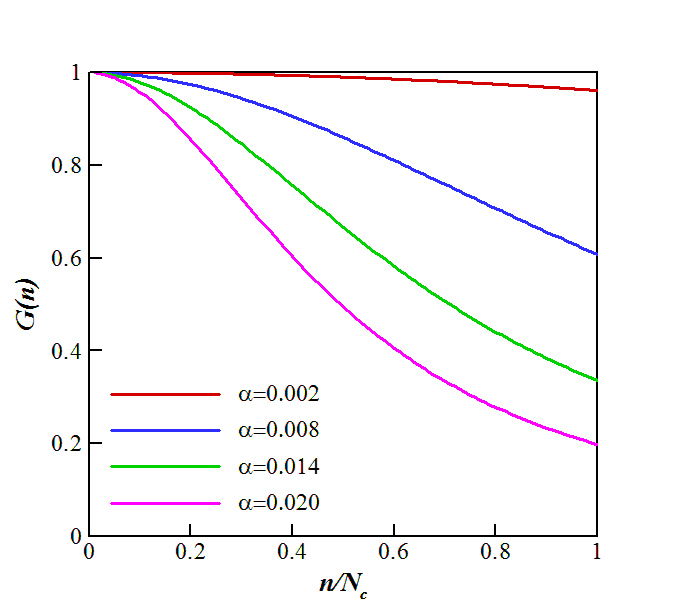}%
 \caption{The differential filter kernel, $G(n)$ for different values of $\alpha$. $N_c$ in this figure is the cutoff wavenumber.}
\label{fig:filter}
\end{center}
\end{figure}
\begin{figure*}[!ht]
\begin{center}
\subfigure[]{
\resizebox*{6.cm}{!}{\includegraphics[trim = 20mm 10mm 20mm 20mm, clip, width=3cm]{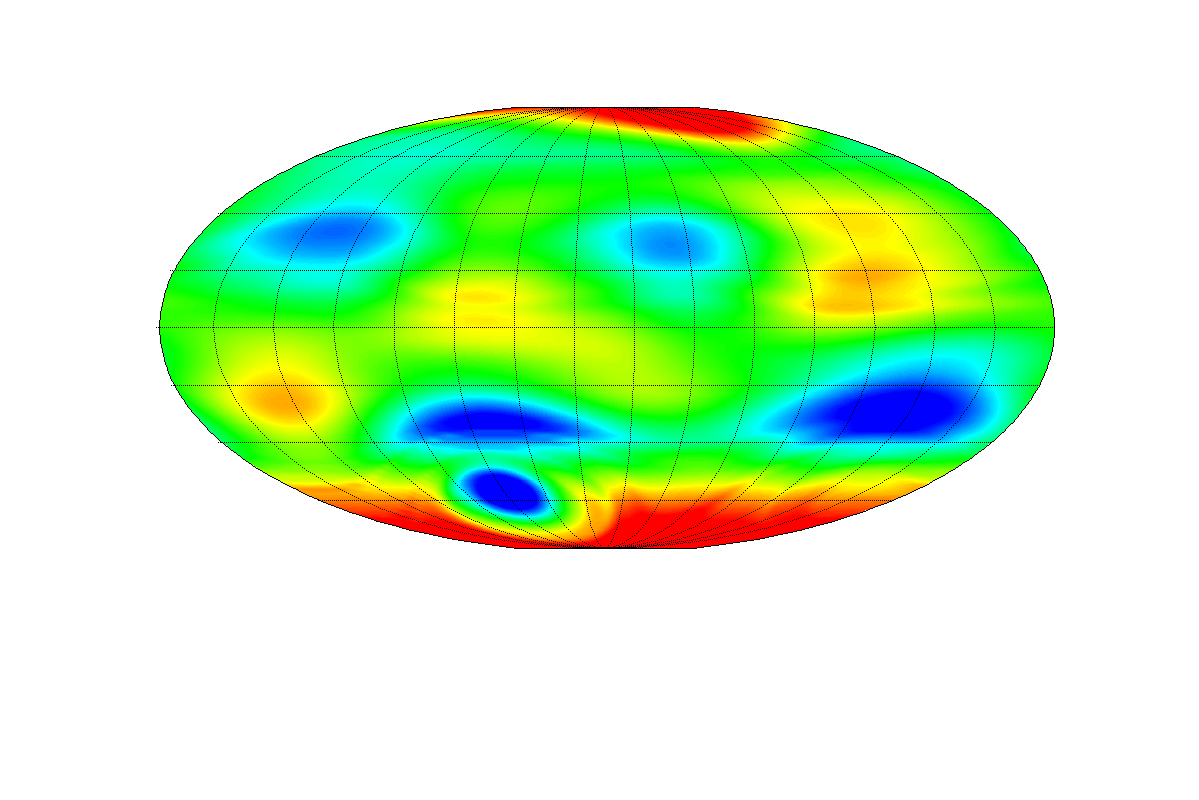}}}%
\subfigure[]{
\resizebox*{6.cm}{!}{\includegraphics[trim = 20mm 10mm 20mm 20mm, clip, width=3cm]{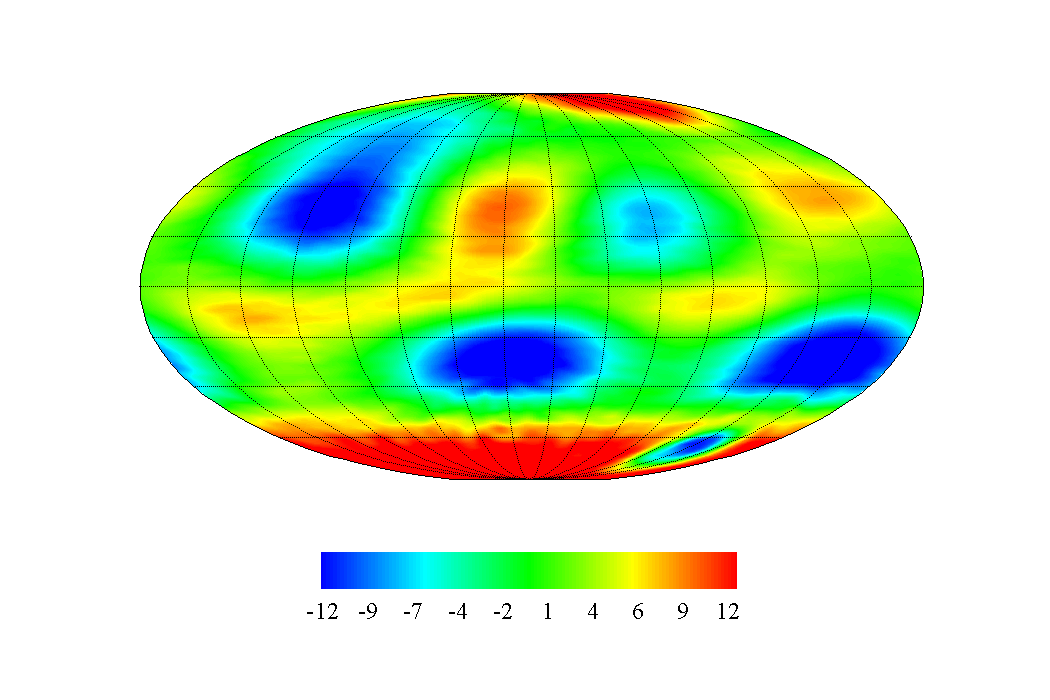}}}%
\subfigure[]{
\resizebox*{6.cm}{!}{\includegraphics[trim = 20mm 10mm 20mm 20mm, clip, width=3cm]{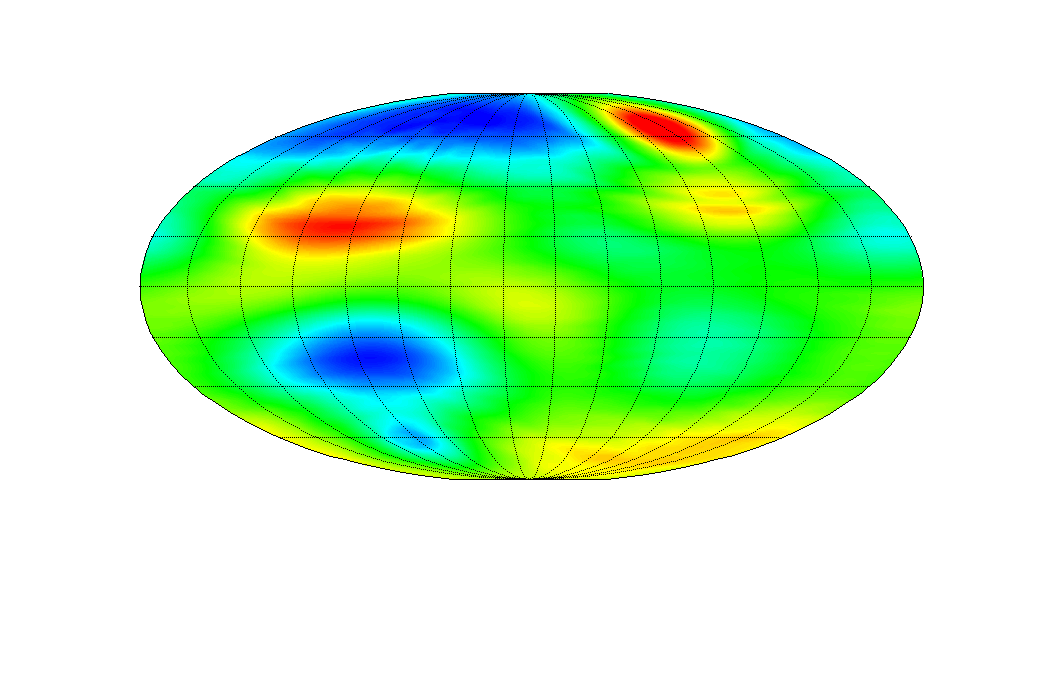}}}%
\caption{The vorticity field for (a) DNS (resolution T333), (b) explicitly filtered LES (resolution T66), and (c) implicitly filtered LES (resolution T66) results for Experiment 1 ($\nu_{2p}=10^{-4}$, $Ro=0.01$).}
\label{fig:v0}
\end{center}
\end{figure*}
\begin{figure*}[!ht]
\begin{center}
\subfigure[]{
\resizebox*{6.cm}{!}{\includegraphics[trim = 20mm 10mm 20mm 20mm, clip, width=3cm]{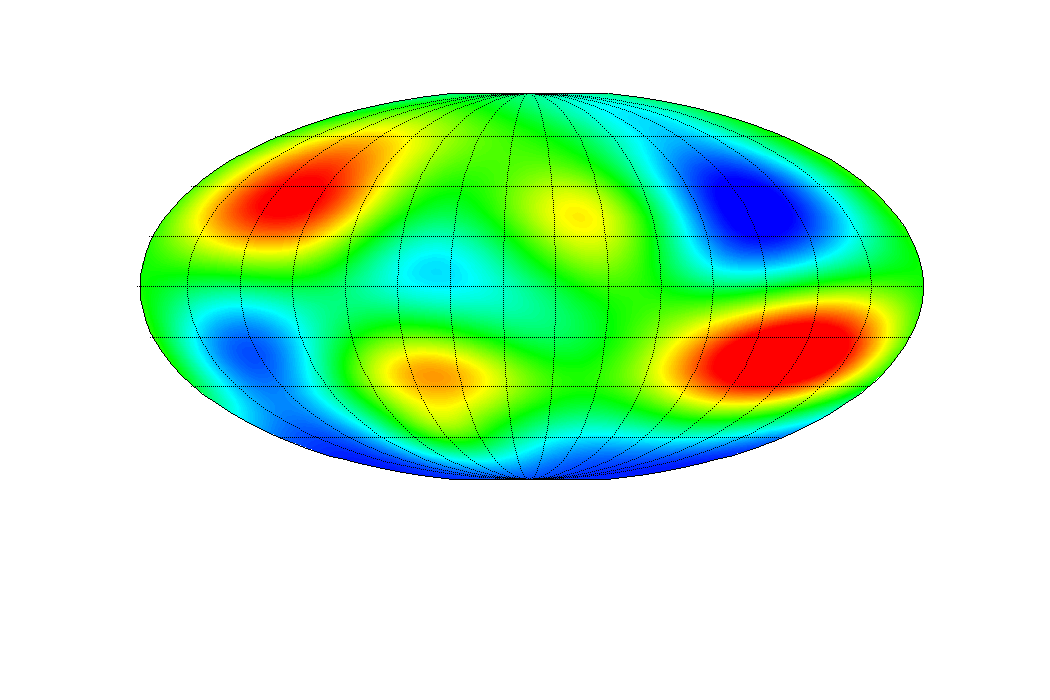}}}%
\subfigure[]{
\resizebox*{6.cm}{!}{\includegraphics[trim = 20mm 10mm 20mm 20mm, clip, width=3cm]{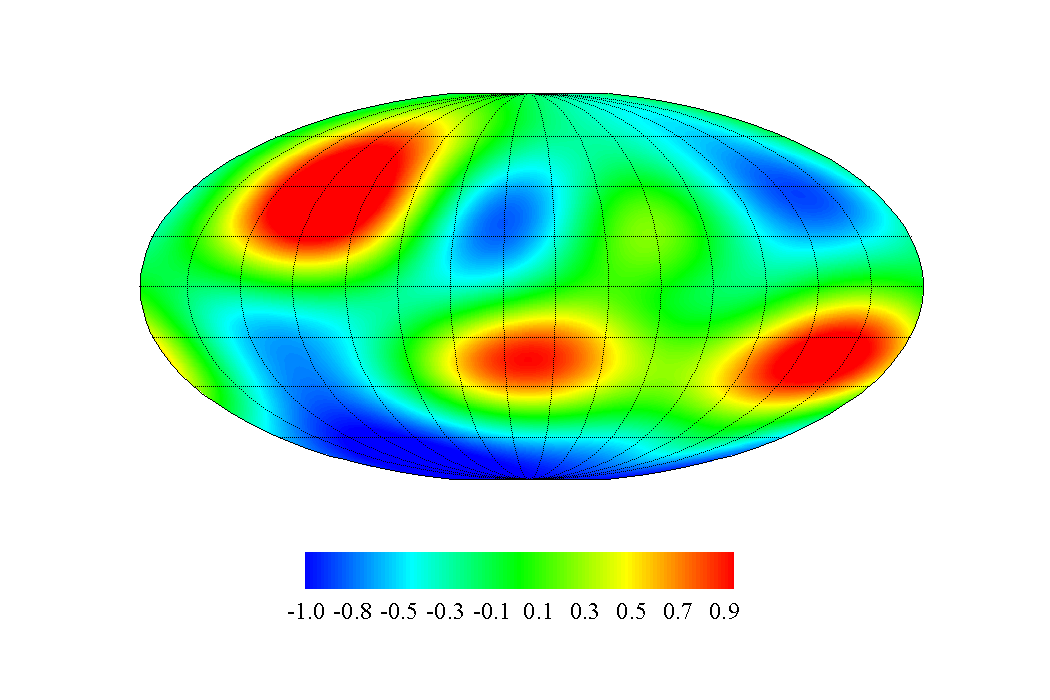}}}%
\subfigure[]{
\resizebox*{6.cm}{!}{\includegraphics[trim = 20mm 10mm 20mm 20mm, clip, width=3cm]{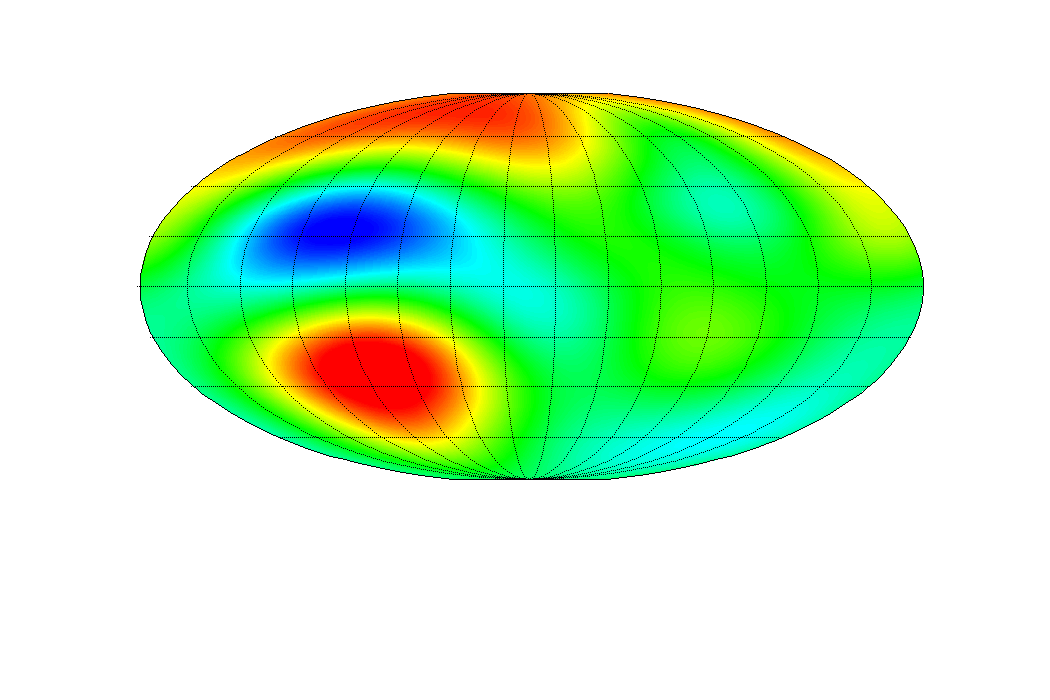}}}%
\caption{The streamfunction field for (a) DNS (resolution T333), (b) explicitly filtered LES (resolution T66), and (c) implicitly filtered LES (resolution T66) results for Experiment 1 ($\nu_{2p}=10^{-4}$, $Ro=0.01$).}
\label{fig:p0}
\end{center}
\end{figure*}

Applying the definition of the Laplacian in spectral space
\begin{equation}
\nabla^2u_{n}^{m}=-\frac{n(n+1)}{R^2}u_{n}^{m}
\label{eq:lap}
\end{equation}
and substituting this for $R=1$ into Eq.~(\ref{eq:difc}), the differential filter kernel in spectral space is defined as
\begin{equation}
G(n)=\frac{1}{1+\alpha n(n+1)}.
\label{eq:sf}
\end{equation}
Plots of $G(n)$ for different values of $\alpha$ are shown in Fig.~\ref{fig:filter}. We use $\alpha =0.005$ in our computations.

Eq.~(\ref{eq:bvef}) is advanced forward in time using a fourth order Runge-Kutta scheme, and the 2/3 dealiasing rule is applied to compute the nonlinear term.

\section{Results and discussion}
\label{sec:res}
In order to investigate the effectiveness of the explicit filtering and the exact deconvolution as a high-level reconstruction model in large eddy simulation of turbulent barotropic flows, we consider two different numerical experiments. 
\subsection{Experiment 1}
For the first experiment we study a flow with a low Rossby number. Furthermore, we apply regular (non-hyper) viscosity,  which corresponds to $p=1$ in Eqs.~(\ref{eq:bve}) and ~(\ref{eq:bvef}). The Rossby number is $Ro=0.01$ and the coefficient of viscosity is $\nu_{2p}= 10^{-4}$. The resolution of the DNS run is $T333$, or $1000\times500$ and for the implicitly and explicitly filtered LES runs the resolution is $T66$, or $200\times100$. In this paper we refer to DNS as the computation with high resolution in which the coherent structures are properly resolved, it does not signify direct numerical simulation in which all the turbulent structures are resolved numerically. In addition, since we do not apply any SGS model, implicitly filtered LES corresponds to an under resolved computation. 

The reconstruction of the subfilter scales in the explicitly filtered LES computation was performed both with the exact deconvolution and the approximate deconvolution of the filtered flow field. Since our results do not show a significant difference between the exact and approximate deconvolution data we present the results obtained by exact reconstruction of the subfilter scales.

The final time of the computation for this experiment is $50$. Contour plots of the vorticity and streamfunction for the DNS, explicitly filtered LES and implicitly filtered LES are shown in Figs.~\ref{fig:v0} and \ref{fig:p0}, respectively. These plots are made at $t=50$ and show the turbulent coherent structures at the final time of computation. Fig.~\ref{fig:v0} and Fig.~\ref{fig:p0} show that the explicitly filtered LES results agree better with the DNS results. The explicitly filtered LES data can predict the correct location of the positive and negative vortices while the implicitly filtered LES results do not produce the correct coherent structures. 
\begin{figure}[!t]
\begin{center}
\includegraphics[trim =1mm 0mm 20mm 20mm, clip,width=0.4\textwidth]{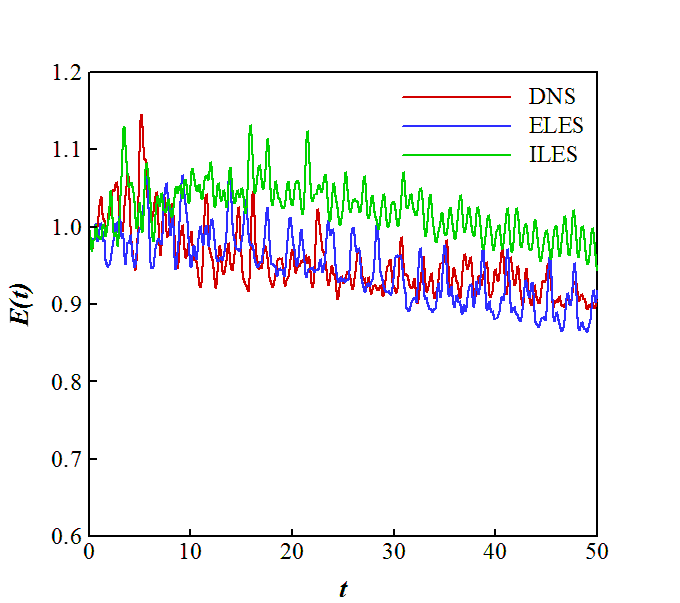}%
\caption{Comparison of the total kinetic energy for the DNS, implicitly filtered LES (ILES), and explicitly filtered LES (ELES) results for Experiment 1 ($\nu_{2p}=10^{-4}$, $Ro=0.01$).}
\label{fig:ke0}
\end{center}
\end{figure}
\begin{figure}[!b]
\begin{center}
\includegraphics[trim =1mm 0mm 20mm 20mm, clip,width=0.4\textwidth]{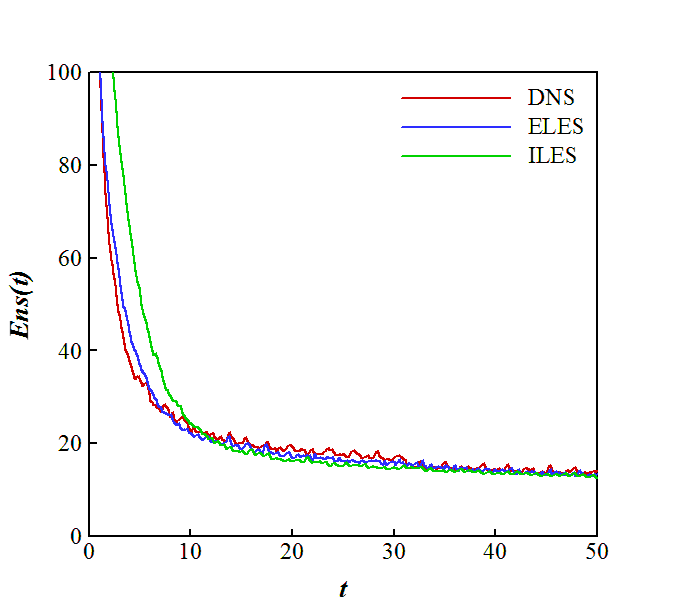}%
\caption{Decay of the total enstrophy for the DNS, implicitly filtered LES (ILES), and explicitly filtered LES (ELES) results for Experiment 1 ($\nu_{2p}=10^{-4}$, $Ro=0.01$).}
\label{fig:ens0}
\end{center}
\end{figure}
\begin{figure}[!t]
\begin{center}
\includegraphics[trim =1mm 0mm 20mm 20mm, clip,width=0.4\textwidth]{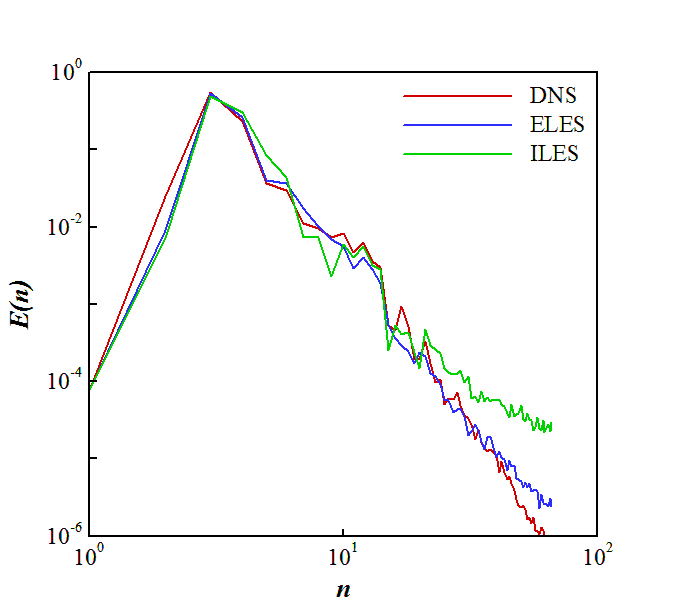}%
\caption{Variation of the energy spectrum with wavenumber for the DNS, implicitly filtered LES (ILES), and explicitly filtered LES (ELES) results for Experiment 1 ($\nu_{2p}=10^{-4}$, $Ro=0.01$).}
\label{fig:es0}
\end{center}
\end{figure}

Variation of the total kinetic energy with time is presented in Fig.~\ref{fig:ke0}. It can be seen that the implicitly filtered LES predicts a higher level of total kinetic energy while the results of the explicitly filtered LES are on top of the DNS results.

Fig.~\ref{fig:ens0} shows the temporal variation of the total enstrophy. Although all three plots converge to the same value at larger times, at the initial time, explicitly filtered LES shows better agreement with the DNS results than the implicitly filtered LES data.

The decay of the energy spectrum with wavenumber in shown in Fig.~\ref{fig:es0}. This plot is made at $t=10$, when the flow has passed the transient state and become fully developed. This figure also shows the effectiveness of the explicit filtering on improving the accuracy of the LES results.

\subsection{Experiment 2}
For the second case we perform the computation at a high Rossby number with a hyperdissipation term. For this experiment the Rossby number is $Ro=1.0$, the constant, $p$, of the hyperdissipation term is $p=8$ and the coefficient of hyperviscosity is $\nu_{2p}=2\times10^{-32}$.  The resolution of the DNS run is T199 which corresponds to $600\times 300$. For implicitly filtered LES and explicitly filtered LES we use the same resolutions as Experiment 1, namely T66. Long-time integration of the BVE at high Rossby numbers produces a vortical quadrupole state \cite{Cho}. To obtain to this quadrupole state we perform the computation until time $t=200$ which is larger than the final time in Experiment 1.
\begin{figure*}[!t]
\begin{center}
\subfigure[]{
\resizebox*{6.cm}{!}{\includegraphics[trim = 20mm 10mm 20mm 20mm, clip, width=3cm]{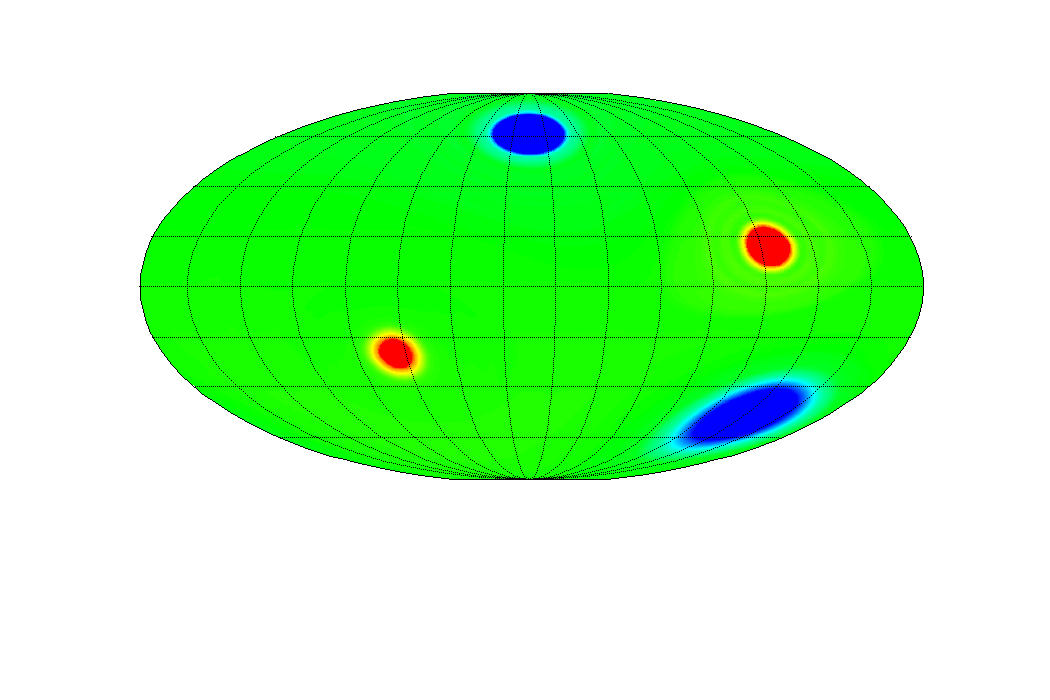}}}%
\subfigure[]{
\resizebox*{6.cm}{!}{\includegraphics[trim = 20mm 10mm 20mm 20mm, clip, width=3cm]{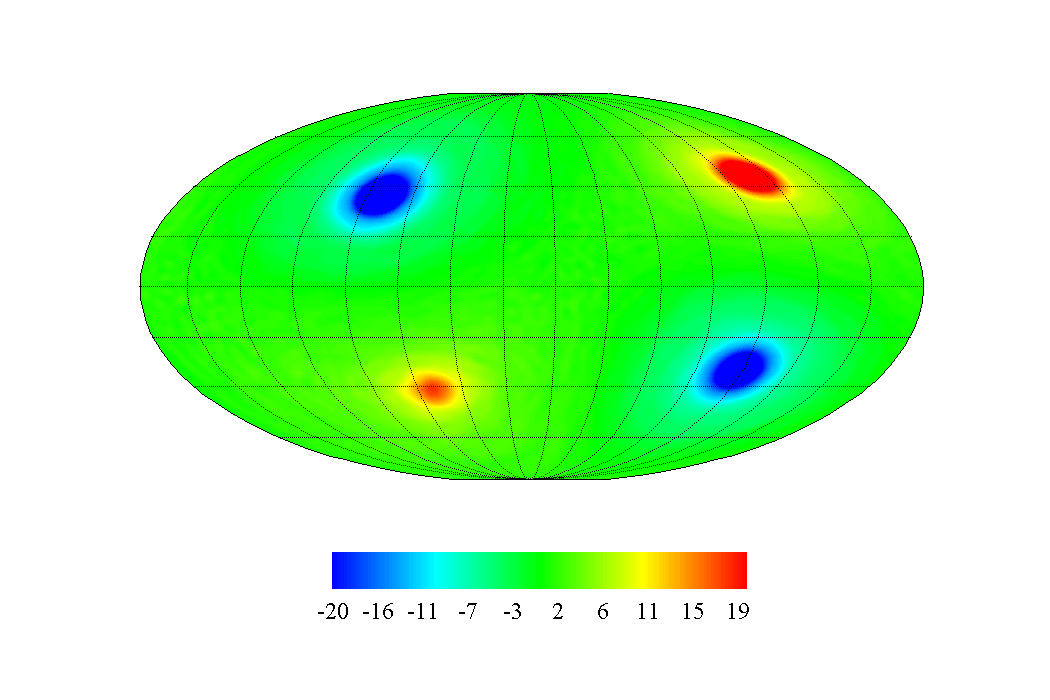}}}%
\subfigure[]{
\resizebox*{6.cm}{!}{\includegraphics[trim = 20mm 10mm 20mm 20mm, clip, width=3cm]{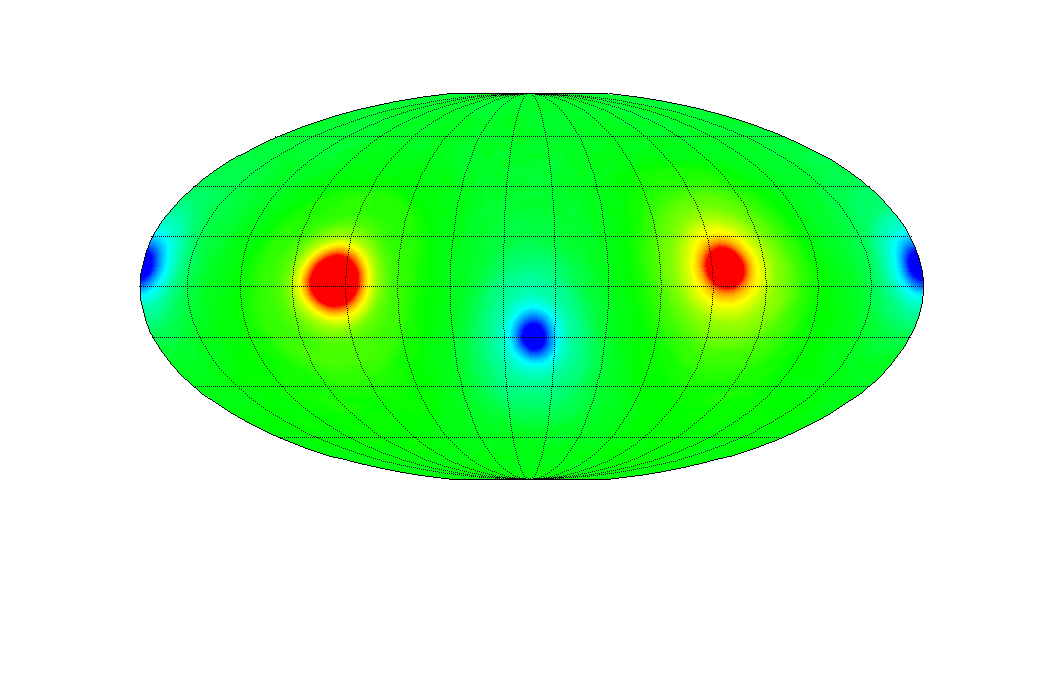}}}%
\caption{The vorticity field for (a) DNS (resolution T199), (b) explicitly filtered LES (resolution T66), and (c) implicitly filtered LES (resolution T66) results for Experiment 2 ($\nu_{2p}=10^{-32}$, $Ro=1.0$).}
\label{fig:v1}
\end{center}
\end{figure*}

\begin{figure*}[!t]
\begin{center}
\subfigure[]{
\resizebox*{6.cm}{!}{\includegraphics[trim = 20mm 10mm 20mm 20mm, clip, width=3cm]{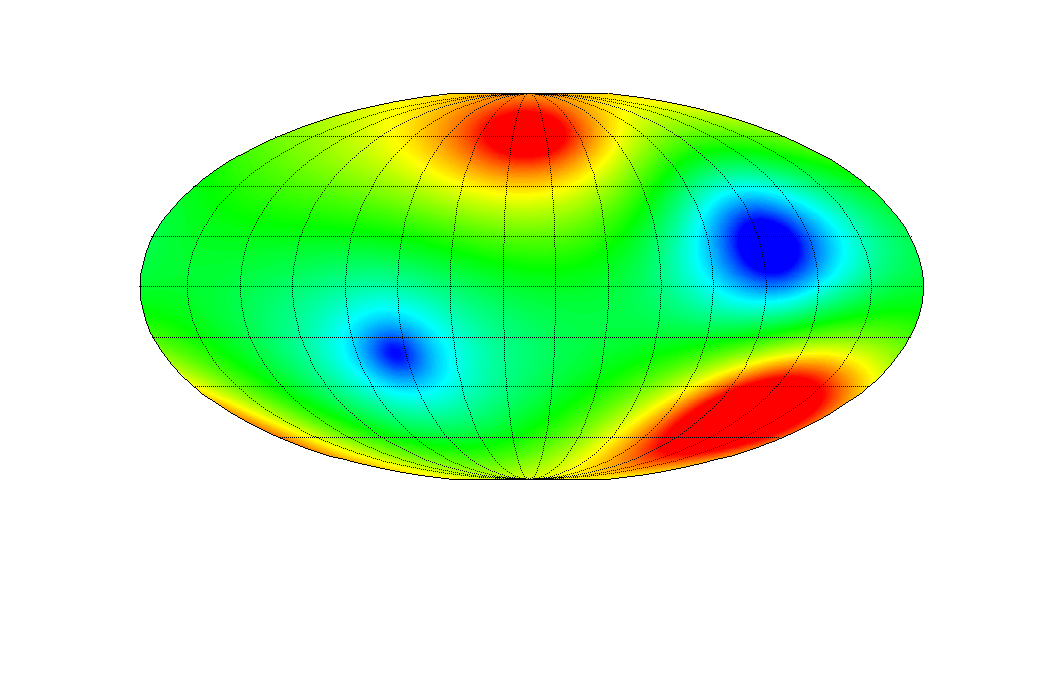}}}%
\subfigure[]{
\resizebox*{6.cm}{!}{\includegraphics[trim = 20mm 10mm 20mm 20mm, clip, width=3cm]{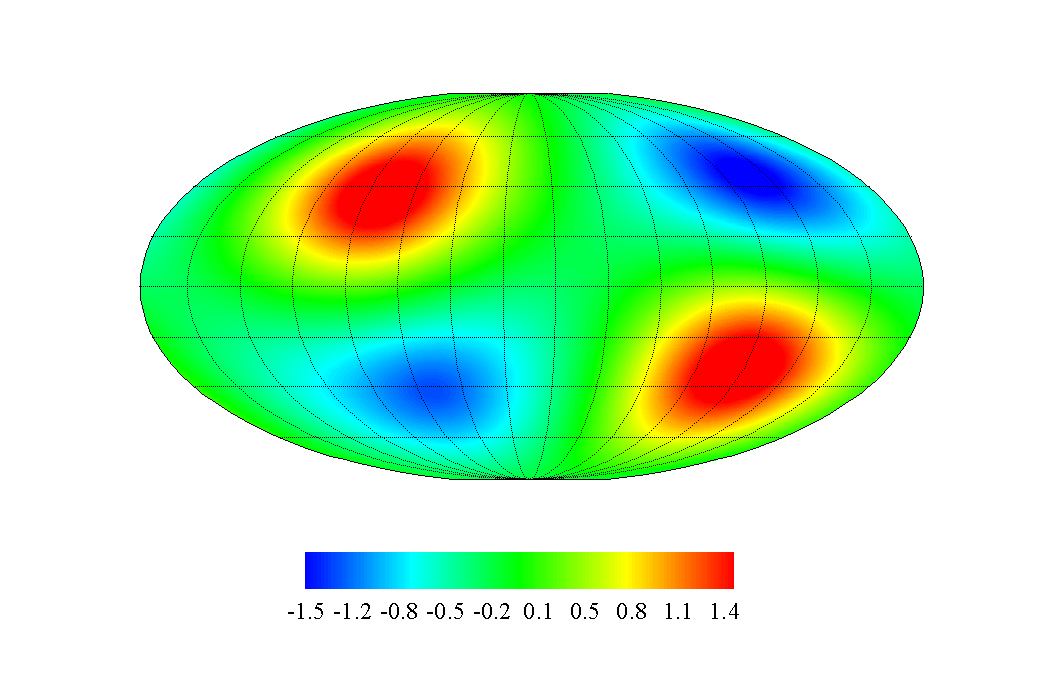}}}%
\subfigure[]{
\resizebox*{6.cm}{!}{\includegraphics[trim = 20mm 10mm 20mm 20mm, clip, width=3cm]{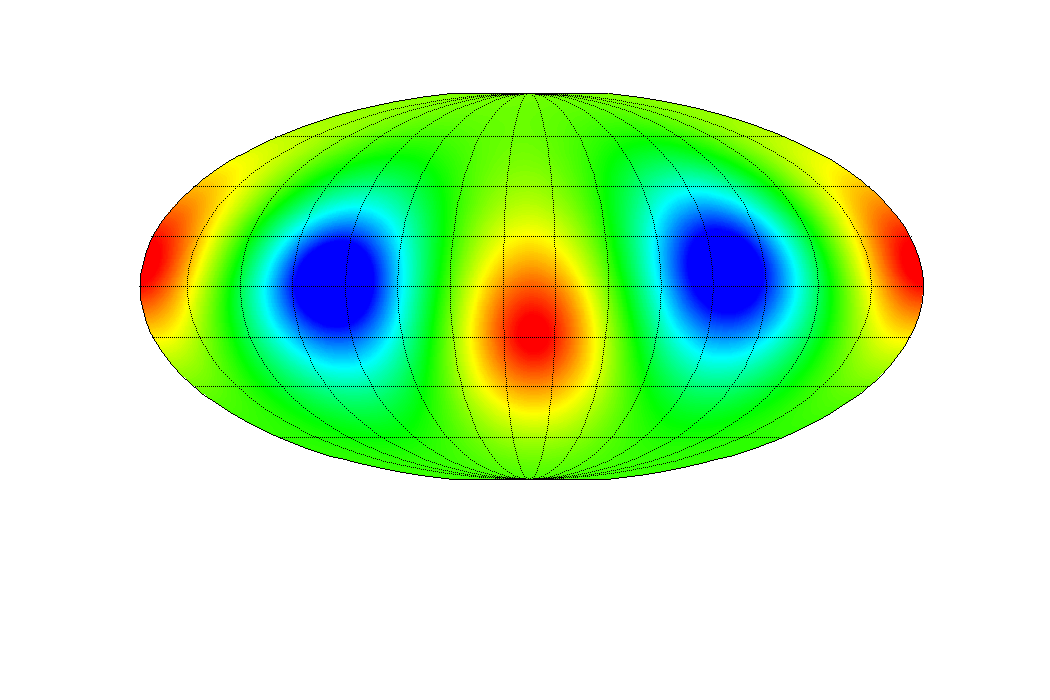}}}%
\caption{The stream function field for (a) DNS (resolution T199), (b) explicitly filtered LES (resolution T66), and (c) implicitly filtered LES (resolution T66) results for Experiment 2 ($\nu_{2p}=10^{-32}$, $Ro=1.0$).}
\label{fig:p1}
\end{center}
\end{figure*}
 \begin{figure}[!b]
\begin{center}
\includegraphics[trim =1mm 0mm 20mm 20mm, clip,width=0.4,width=0.4\textwidth]{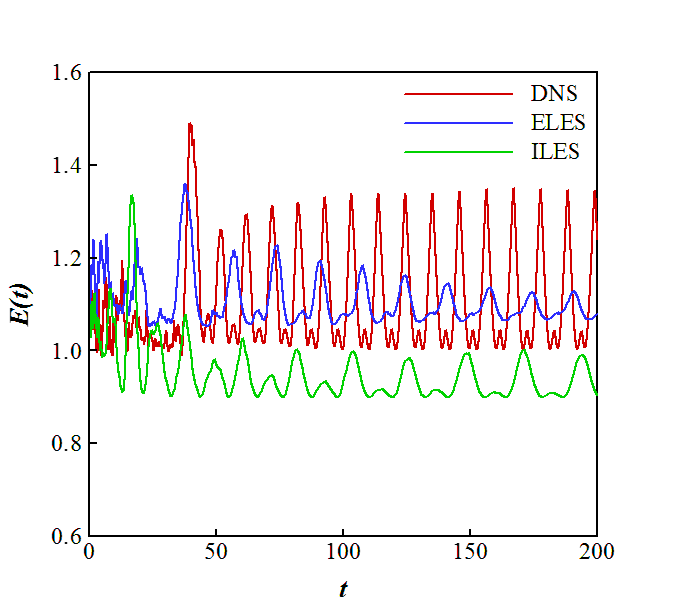}%
\caption{Comparison of the total kinetic energy for the DNS, implicitly filtered LES (ILES), and explicitly filtered LES (ELES) results for Experiment 2 ($\nu_{2p}=10^{-32}$, $Ro=1.0$).}
\label{fig:ke1}
\end{center}
\end{figure}

\begin{figure}[!b]
\begin{center}
\includegraphics[trim =1mm 0mm 20mm 20mm, clip,width=0.4,width=0.4\textwidth]{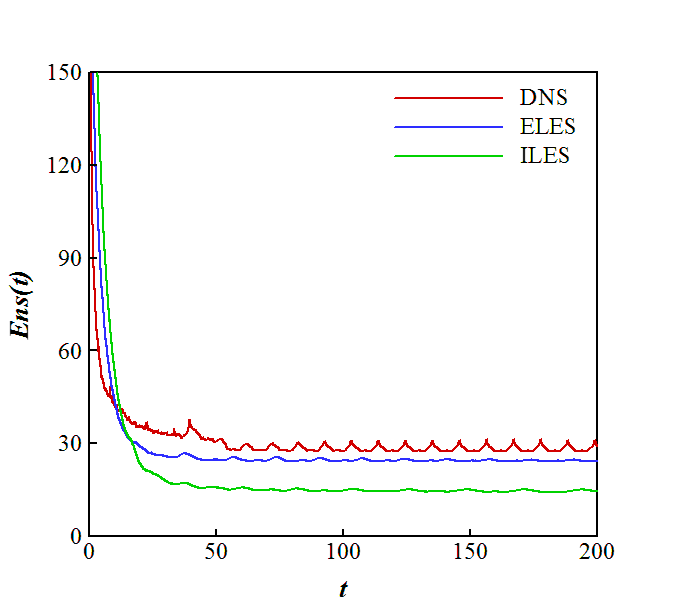}%
\caption{Decay of the total enstrophy for the DNS, implicitly filtered LES (ILES), and explicitly filtered LES (ELES) results for Experiment 2 ($\nu_{2p}=10^{-32}$, $Ro=1.0$).}
\label{fig:ens1}
\end{center}
\end{figure}

\begin{figure}[!t]
\begin{center}
\includegraphics[trim =1mm 0mm 20mm 20mm, clip,width=0.4,width=0.4\textwidth]{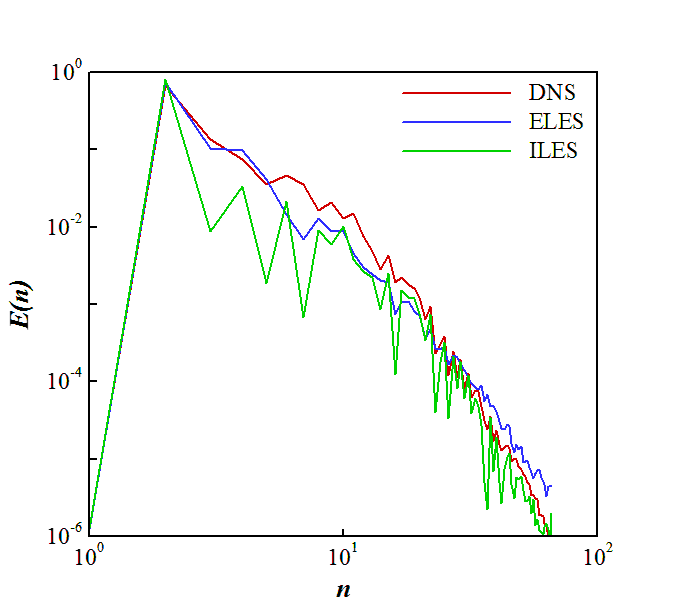}%
\caption{Variation of the energy spectrum with wavenumber for the DNS, implicitly filtered LES (ILES), and explicitly filtered LES (ELES) results for Experiment 2 ($\nu_{2p}=10^{-32}$, $Ro=1.0$).}
\label{fig:es1}
\end{center}
\end{figure}

Similar to Experiment 1, the results presented here for the explicitly filtered LES computation are obtained by exact reconstruction of the subfilter scales.

Vorticity and streamfunction fields are presented in Figs.~\ref{fig:v1} and \ref{fig:p1}, respectively. All plots show the formation of the quadrupole structure at the final time of the computation. As in Experiment 1, the results of the explicitly filtered LES predict the location of the coherent structures better than the implicitly filtered LES results.
 
The temporal variation of the total kinetic energy is shown in Fig.~\ref{fig:ke1}. Interestingly, at high Rossby numbers the total kinetic energy, computed from the implicitly filtered LES, is at a lower level than the total kinetic energy from the DNS results. It seems that performing explicit filtering in LES injects energy via backscatter, increases the energy level, and shows better agreement with the DNS results.

Variation of the total enstrophy with time is shown in Fig.~\ref{fig:ens1}. This figure shows that in contradiction to Experiment 1, the DNS, explicitly filtered LES and implicitly filtered LES results converge to different values. However, the explicitly filtered LES results are in better agreement with the DNS results.  

Fig.~\ref{fig:es1} shows the variation of the energy spectrum with wavenumber at $t=50$ when the flow is fully developed. It can be seen that the explicitly filtered LES results show better agreement with the DNS data at low wavenumbers while at higher wavenumbers, implicitly filtered LES shows better agreement. 

In all comparisons, the explicitly filtered LES results show a better match with the DNS results.

\section{Conclusions}
\label{sec:con}
We investigated the effect of explicit filtering on the accuracy of large eddy simulations results of the turbulent barotropic vorticity equation (BVE) on the sphere as a first model of Earth's atmosphere. We used a differential filter with a constant width and the unfiltered flow variables were exactly reconstructed by dividing the filtered flow variables by the filter kernel. In order to study the pure effects of the explicit filtering alone, no subgrid scale (SGS) closure model was applied. Two main features distinguish our work from previous studies: (1) The computations were performed in spherical coordinates. The spherical geometry is important in computations of large scale geophysical motions. However, due to the complexity of computations in spherical coordinates, few studies have been performed on explicit filtering in LES of turbulent flows in spherical coordinates. (2) We performed explicit filtering in LES of the barotropic flow using a spectral method based on spherical harmonic transforms.  Previous studies \cite{Winckelmans2} have shown that explicit filtering in spectral simulations of isotropic turbulence based on Fourier transforms does not show improvement over implicit filtering, while spectral computations of turbulent channel flow based on Fourier-Chebyshev transforms showed significant improvement in the explicit filtering results over implicit filtering results \cite{Domaradzki,Stolz2}. The reasons why explicit filtering did not work in spectral simulations of isotropic turbulence but worked in spectral simulations of turbulent channel flow may be the difference between the spectral methods or/and the difference between the reconstruction models used in each case. Here, we investigated the effect of the reconstruction model by applying exact deconvolution and approximate deconvolution with $N=3$. Our results showed that although there are small differences between the results obtained from exact deconvolution and the results obtained from approximate deconvolution, both models showed significant improvement in explicitly filtered LES results compare to implicitly filtered LES results. It can be seen that the reconstruction model has only a small influence on the effectiveness of explicit filtering and the main reason why explicit filtering improves the results in spectral computations of channel flow and BVE is related to Fourier-Chebyshev and Fourier-Legendre basis functions. Here, we performed two different experiments. In the first experiment we applied the standard viscosity and studied a flow with a low Rossby number. In the second case we used hyperviscosity and investigated the flow with a large Rossby number. In both experiments contour plots of vorticity and streamfunction for implicitly filtered LES and explicitly filtered LES were compared with the DNS results. It was shown that explicitly filtered LES can predict the behavior of the coherent structures more accurately than implicitly filtered LES. Temporal variations of the total kinetic energy and total enstrophy with time and decay of the energy spectrum with wavenumber also showed the superior performance of the explicit filtering compare to implicit filtering. In general, we conclude that for LES of turbulent flows in spectral space using spherical harmonic expansions, explicitly filtered LES results show much better agreement with DNS results than implicitly filtered LES results, and that the effects of SGS models should also be considered, which will be the subject of a future study.

\nocite{*}
\bibliography{Manuscript}

\end{document}